\documentclass[journal]{IEEEtran}

\usepackage{mathtools,amssymb,lipsum, nccmath}

\usepackage{cuted}
\usepackage{cite}

\usepackage{amsmath,amssymb,amsfonts}
\usepackage{mathtools}
\usepackage[linesnumbered,ruled,vlined]{algorithm2e}
\usepackage{algcompatible}

\SetCommentSty{mycommfont}
\usepackage[flushleft]{threeparttable}
\usepackage{footnote}
\usepackage{nicematrix,enumitem,booktabs}
\makesavenoteenv{tabular}
\usepackage{xcolor}
\usepackage{color}
\usepackage{url}
\usepackage{mdwmath}
\usepackage{mdwtab}
\usepackage[all]{xy}
\usepackage{array}
\usepackage[tight,footnotesize]{subfigure}
\usepackage{stfloats}
\usepackage{bbm} 
\usepackage{adjustbox}
\usepackage{academicons}
\newcommand{\orcid}[1]{\href{https://orcid.org/#1}{\textcolor[HTML]{A6CE39}{\aiOrcid}}}
\usepackage{graphicx}
\DeclareGraphicsExtensions{.pdf,.jpeg,.eps}
\usepackage{multicol}
\usepackage{graphicx}
\usepackage{setspace}
\newcommand\scalemath[2]{\scalebox{#1}{\mbox{\ensuremath{\displaystyle #2}}}}

\SetKwInput{KwData}{Input}
\SetKwInput{KwResult}{Output}

\begin{document}
	\title{A Probabilistic Shaping Approach for\\ 
		Optical Region-of-Interest Signaling}
	
	\author{Duc-Phuc~Nguyen,~\IEEEmembership{Member,~IEEE,}
		Yoshifumi~Shiraki,~\IEEEmembership{Member,~IEEE,}
		Jun~Muramatsu,~\IEEEmembership{Senior Member,~IEEE,}
		and~Takehiro~Moriya,~\IEEEmembership{Fellow,~IEEE}
		\thanks{Manuscript received December 2, 2021; revised January 20, 2022; accepted February 15, 2022. (\emph{Corresponding author: Duc-Phuc Nguyen}).}
		\thanks{The authors are with NTT Communication Science Laboratories, Nippon Telegraph and Telephone Corporation. 3-1, Morinosato-Wakamiya, Atsugi, Kanagawa, 243-0198, Japan (e-mail: nguyenducphuc@ieee.org).}
	   }
	
	\markboth{}%
	{Shell \MakeLowercase{\textit{et al.}}: Bare Demo of IEEEtran.cls for IEEE Journals}
	
	\maketitle
	
	\begin{abstract}
		We propose a probabilistic shaping approach for region-of-interest  signaling, where a low-rate signal controls the desired probabilistic ranges of a high-rate data stream using a flexible distribution controller. In addition, we introduce run-length-aware values for frozen bit indices in systematic polar code to minimize the run-length without using run-length-limited code. Our compact system can support soft-decision forward-error-correction decoding with excellent spectral efficiency compared with related work based on hybrid modulation schemes. 
	\end{abstract}
	
	
	\IEEEpeerreviewmaketitle
	
	\section{Introduction}
	\label{intro}
	\IEEEPARstart{R}{egion}-of-interest (RoI) signaling has been introduced by IEEE 802.15.7m group (TG7m) for vehicular optical camera communication (OCC) systems \cite{thieu2019optical}. In this technique, the low-rate and high-rate data streams are transmitted simultaneously. The low-rate stream is used for RoI identification, whereas the high-rate stream is used for high-speed data communication on the selected RoI. Low-rate and high-rate streams are modulated by hybrid modulation schemes such as twinkle variable pulse position modulation (VPPM), or hybrid spatial phase-shift keying (HS-PSK). In detail, the low-rate stream is modulated by undersampled frequency-shift on-off keying (UFSOOK), undersampled phase-shift on-off keying (UPSOOK), or spatial-2 phase-shift keying (S2-PSK), whereas the high-rate stream is modulated by twinkle VPPM or dimming spatial-8 phase-shift keying (DS8-PSK) \cite{nguyen2018technical}. Both streams can be decoded by a dual-camera receiver, where the low-speed camera decodes the low-rate stream, and a high-speed camera decodes the high-rate stream. Hence, the computational load of the single-camera receiver is mitigated. 
	
	Dimming support is one of the critical considerations in visible light communication (VLC) and OCC specifications. Dimming can be controlled by adjusting the ON (1's) or OFF (0's) ratio, which varies from 0\% to 100\%. In normal cases, dimming control comprises puncturing and compensation symbol (CS) insertion where many redundant bits are added to the frame, which degrades the achievable rate \cite{wang2017dimming}. In addition, the dimming method in twinkle VPPM and HS-PSK changes the position and duty cycle of groups of high-rate pulses over time. Hence, it is challenging to demodulate twinkle VPPM and DS8-PSK signals on the receiver side because each pulse must be sampled, synchronized, and decoded at rigorous timing \cite{noh2015dimming}. In conventional VLC systems, run-length-limited (RLL) codes keep DC balance with equal 1's and 0's in every symbol to avoid long runs of 1's and 0's, which causes flicker and clock and data recovery (CDR) issues. In fact, there is a trade-off between ease of implementation and rate loss in RLL codes such as Manchester, 4B6B, 8B10B \cite{rajagopal2012ieee}. Besides, most RLL decoders produce hard outputs for hard-decision (HD) forward error correction (FEC) decoder in joint FEC-RLL based systems, which show poor error-correction performance. A soft-input soft-output (SISO) RLL decoder has been introduced by Kim et al. \cite{wang2015soft}, which produces soft outputs for soft-decision (SD) FEC decoding algorithms to improve the overall system reliability. However, the decoding algorithm of the SISO RLL code is bulky and hard to adopt in VLC receivers. 
	
	In this letter, we propose a probabilistic shaping approach for RoI signaling applications based on a flexible distribution controller (FDC) and systematic polar encoder/decoder (SPE/SPD) with run-length-aware (RLA) frozen bit values. The proposed system can simultaneously transmit high-rate and low-rate data with low rate-loss overhead. Besides supporting SD FEC decoders, our system guarantees the flicker mitigation without using RLL code.

	\begin{figure}[!t]
		\centerline{\includegraphics[width=1\columnwidth]{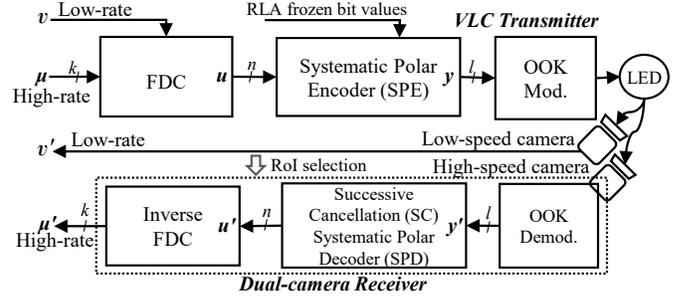}}
		\caption{The proposed system}
		\label{fig_1}
	\end{figure}
	
	\section{System Model}
	\subsection{Proposed system} 
	\label{proposed_system}
	Fig. \ref{fig_1} shows the block diagram of the proposed system. The dual inputs of the system are a high-rate binary sequence $\mu = (\mu_1,\mu_2,...,\mu_k)$ and a low-rate signal $v$ for $v=0,1$. We denote the ratios of 1’s in a $n$-bit FDC codeword $u$ and a $l$-bit FEC codeword $y$ as $\phi(\%)$ and $\omega(\%)$, respectively. In our transmitter, the FDC encodes a binary vector $\mu$ with Bernoulli $(\frac{1}{2})$ distribution into a codeword $u = (u_1,u_2,...,u_n)$ with a desired distribution. Indeed, $\omega$ is different from $\phi$ when the message $u$ is encoded by a FEC encoder. We apply an SPE to preserve probabilistic shapes of $\omega$ for the two dimming ranges. In addition, RLA frozen bit values are applied to the SPE to limit the run-length of $y$, instead of using RLL code as in conventional VLC systems. For a dual-camera receiver, a low-speed camera (frame rate $<$100 frames per second ($fps$)) with long exposure time detects the dimming ranges of light-emitting diode (LED) and decodes the low-rate bit sequence, which is used for RoI identification. A high-speed global shutter camera (frame rate $>$ 10000 $fps$) decodes the high-rate data stream on the selected RoI connection. An SD successive cancellation SPD (SD-SPD) decodes the codeword $y'$ for $u'$. Finally, the inverse FDC decodes the message $u'$ for high-rate binary sequence $\mu'$. OOK modulation and demodulation are selected because of their simplicity and popularity.
	
	\subsection{Flexible distribution controller (FDC)} 
	\label{subsection_FDC}
	 The distribution controller (DC) modifies the probability of 1's in a sequence of bits to a desired distribution using arithmetic decompression algorithm. Let $S' = \{ {u_1},{u_2},...,{u_{M - 1}},{u_M}\} $ where $M=\smash{(\scalemath{0.7}{\begin{matrix}n \\m \\\end{matrix}})}$ denote the set of binary $n$-tuples with weight $m$. A source model which emits $M$ sequences in $S'$ with probability equal to $P({u_i}) = 1/M$ for $i = 1,2,...,M - 1,M$. Other sequences of length $n$ which are not in $S'$ have zero probability. Let $S = \{ {\mu _1},{\mu _2},...,{\mu _K}\} $ where $K \le M$ denote the set of $k$-bit information words. A one-to-one correspondence is established between $K$ information words in $S$ and a subset of $K$ of $M$ sequences in $S'$  based on the source statistics introduced in \cite{ramabadran1990coding}. The code-rate loss $\Delta_{DC}$, which indicates the average number of redundant bits carried by each DC-encoded bit, is given as 
	
\begin{equation}\label{rateloss}
	\resizebox{0.8\hsize}{!}{$%
    	{\Delta _{DC}} = 1 - \frac{{\left\lfloor {{{\log }_2}\left( {\begin{array}{*{20}{c}}
						n\\
						m
				\end{array}} \right)} \right\rfloor }}{n} = 1 - \frac{{\left\lfloor {{{\log }_2}\left( {\begin{array}{*{20}{c}}
						n\\
						{n - m}
				\end{array}} \right)} \right\rfloor }}{n}
$%
}%
\end{equation}
	where $\lfloor...\rfloor$ denotes the floor function, $\lfloor \log_2\smash{(\scalemath{0.8}{\begin{matrix}n \\m \\\end{matrix}})}\rfloor$ outputs the largest $k$ satisfying the inequality $K \le M$ where $K = {2^k}$.  
		
\begin{figure}[!t]
	\centerline{\includegraphics[width=0.95\columnwidth]{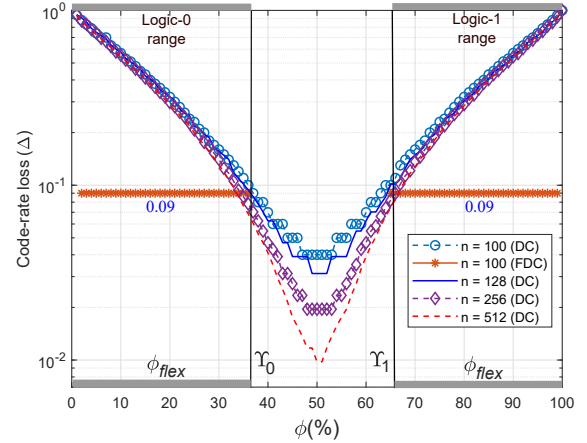}}
	\caption{Rate loss of DC and FDC at different code lengths. Logic-0 probabilistic range is defined by $\phi_{flex}\in(0\%, \Upsilon_0]$ where $\Upsilon_0=36\%$. Logic-1 probabilistic range is defined by $\phi_{flex}\in[\Upsilon_1, 100\%)$ where $\Upsilon_1=64\%$.}
	\label{fig2}
\end{figure}

	Fig. \ref{fig2} shows the rate-loss analysis for various DC code lengths. For code length $n=100$, it can be seen that $\Delta_{DC} > 0.1$ when $\phi<36\%$ and $\phi>64\%$. Hence, we select 0.1 as a maximum threshold of $\Delta_{DC}$ to design probabilistic ranges of DC outputs. We denote the maximum $\phi$ of the logic-0 probabilistic range as $\Upsilon_0$ $(0\%, \Upsilon_0]$ and  the minimum $\phi$ of the logic-1 probabilistic range as $\Upsilon_1$ $[\Upsilon_1, 100\%)$, where $\Delta_{DC} = 0.1$ when $\phi={\Upsilon_0,\Upsilon_1}$. For $n=100$, $\Upsilon_0 = 36\%$ and $\Upsilon_1 = 64\%$. The reason for $\Upsilon_0\notin(36\%, 50\%]$ and $\Upsilon_1\notin[50\%, 64\%)$ is that the low $fps$ camera on the receiver cannot discriminate the LED ON or OFF states when $\phi$ is close to $50\%$. In addition, it can be seen that $\Upsilon_0<36\%$ and $\Upsilon_1>64\%$ when $n>100$. However, we select $n=100$ and use FEC code length $l$ = 128, which is compatible with short frame lengths in most VLC specifications. 
	
	We propose the FDC, which produces codewords with a flexible distribution of $\phi$, which is denoted as $\phi_{flex}$. The $\phi_{flex}$ can be any value in the logic-0 probabilistic range ($0\%, \Upsilon_0$] or logic-1 probabilistic range [$\Upsilon_1, 100\%$). A low-rate signal $v$ controls which probabilistic range the $\phi_{flex}$ stays in. For instance, if $v=0$, the FDC will produce a codeword $u$, which has $\phi_{flex}$ in the range $(0\%, \Upsilon_0]$; if $v=1$,  it will produce one with $\phi_{flex}$ in the range $[\Upsilon_1, 100\%)$. For $1\leq m<\frac{n}{2}$, let ${C}_{m}$ denote how many times larger the codebook of $n$-tuples of weight $m$ is than the codebook of $n$-tuples of weight $m-1$, which is given as  
		
\begin{align}	
	\label{eq1}	
	\begin{split}
		{{C}_{m}=\left( \begin{matrix}
				n  \\
				m  \\
			\end{matrix} \right)}/{\left( \begin{matrix}
				n  \\
				m-1  \\
			\end{matrix} \right)}\;\ =\left( \frac{n-m+1}{m} \right)
	\end{split}
\end{align}
The number of codewords in the codebook of the FDC when $\phi_{flex}\in(0\%, \Upsilon_0]$ and $\phi_{flex}$ $\in$ $[\Upsilon_1, 100\%)$ are given as follow
			\begin{equation}\label{eq2}
		\resizebox{0.9\hsize}{!}{$%
			{Z_{(0,{\Upsilon _0}]}} = \left( {\begin{array}{*{20}{c}}
					n\\
					1
			\end{array}} \right) + \left( {\begin{array}{*{20}{c}}
					n\\
					2
			\end{array}} \right) + ... + \left( {\begin{array}{*{20}{c}}
					n\\
					{{\Upsilon _0}n - 1}
			\end{array}} \right) + \left( {\begin{array}{*{20}{c}}
					n\\
					{{\Upsilon _0}n}
			\end{array}} \right)
			$%
		}%
	\end{equation}

\begin{equation}\label{eq3}
\resizebox{0.9\hsize}{!}{$%
{Z_{[{\Upsilon _1},1)}} = \left( {\begin{array}{*{20}{c}}
		n\\
		{{\Upsilon _1}n}
\end{array}} \right) + \left( {\begin{array}{*{20}{c}}
		n\\
		{{\Upsilon _1}n + 1}
\end{array}} \right) + ... + \left( {\begin{array}{*{20}{c}}
		n\\
		{n - 2}
\end{array}} \right) + \left( {\begin{array}{*{20}{c}}
		n\\
		{n - 1}
\end{array}} \right)
	$%
}%
\end{equation}
Assume that ${\Upsilon _0} = 1 - {\Upsilon _1}$, we have

\begin{equation}\label{eq4}
	\resizebox{0.8\hsize}{!}{$%
\left( {\begin{array}{*{20}{c}}
		n\\
		{{\Upsilon _1}n}
\end{array}} \right) = \left( {\begin{array}{*{20}{c}}
		n\\
		{(1 - {\Upsilon _0})n}
\end{array}} \right) = \left( {\begin{array}{*{20}{c}}
		n\\
		{n - {\Upsilon _0}n}
\end{array}} \right) = \left( {\begin{array}{*{20}{c}}
		n\\
		{{\Upsilon _0}n}
\end{array}} \right)
$%
}%
\end{equation}

\begin{equation}\label{eq5}
	\resizebox{0.9\hsize}{!}{$%
\left( {\begin{array}{*{20}{c}}
		n\\
		{{\Upsilon _1}n + 1}
\end{array}} \right) = \left( {\begin{array}{*{20}{c}}
		n\\
		{{\Upsilon _0}n - 1}
\end{array}} \right);\begin{array}{*{20}{c}}
	{}& \ldots &{}
\end{array};\left( {\begin{array}{*{20}{c}}
		n\\
		{n - 1}
\end{array}} \right) = \left( {\begin{array}{*{20}{c}}
		n\\
		1
\end{array}} \right)
$%
}%
\end{equation}
From (\ref{eq1})(\ref{eq4})(\ref{eq5}), (\ref{eq2})(\ref{eq3}) can now be written as

\begin{equation}\label{eq6}
\resizebox{0.9\hsize}{!}{$%
{Z_{(0,{\Upsilon _0}]}} = {Z_{[{\Upsilon _1},1)}} = \left( {\begin{array}{*{20}{c}}
		n\\
		1
\end{array}} \right) + {C_2}\left( {\begin{array}{*{20}{c}}
		n\\
		1
\end{array}} \right) + ... + {C_{{\Upsilon _0}n - 1}}\left( {\begin{array}{*{20}{c}}
		n\\
		1
\end{array}} \right) + {C_{{\Upsilon _0}n}}\left( {\begin{array}{*{20}{c}}
		n\\
		1
\end{array}} \right)
$%
}%
\end{equation}
To calculate the code-rate loss of the FDC, which is denoted as ${\Delta_{FDC}}$, from $\Delta_{DC}$, we replace  $\smash{(\scalemath{0.7}{\begin{matrix}n \\m \\\end{matrix}})}$ in  (\ref{rateloss}) with (\ref{eq6}), we have

\begin{equation}\label{eq7}
	\resizebox{0.9\hsize}{!}{$%
\begin{array}{l}
	{\Delta _{FDC(0,{\Upsilon _0}]}} = {\Delta _{FDC[{\Upsilon _1},1)}} = 1 - \frac{{\left\lfloor {{{\log }_2}\left( {{Z_{(0,{\Upsilon _0}]}}} \right)} \right\rfloor }}{n}\\
	= 1 - \frac{{\left\lfloor {{{\log }_2}\left( {(1 + {C_2} + ... + {C_{{\Upsilon _0}n - 1}} + {C_{{\Upsilon _0}n}})\left( {\begin{array}{*{20}{c}}
							n\\
							1
					\end{array}} \right)} \right)} \right\rfloor }}{n}
\end{array}
$%
}%
\end{equation}
From (\ref{eq7}), it can be seen that ${\Delta _{FDC(0,{\Upsilon _0}]}}$ and ${\Delta _{FDC[{\Upsilon _1},1)}}$ are symmetric with respect to $\phi=50\%$ when ${\Upsilon _0} = 1 - {\Upsilon _1}$.

\subsection{RLA frozen bit values for SPC}
	
	\begin{figure}[!t]
		\centerline{\includegraphics[width=0.95\columnwidth]{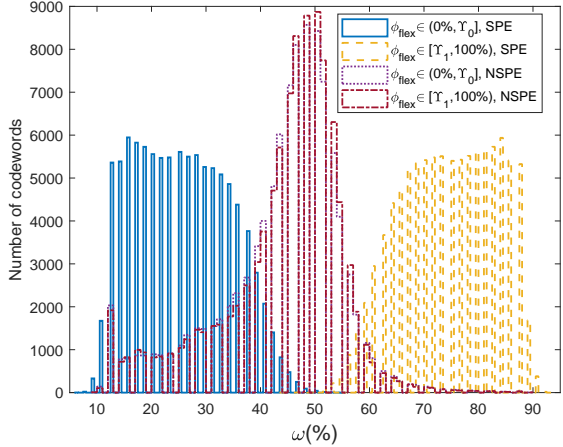}}
		\caption{Probabilistic distribution of $\omega$ in 100,000 codewords encoded by the SPE and NSPE. To create two dimming ranges, the FDC produces outputs with $\phi_{flex}\in(0\%, \Upsilon_0]$ and $\phi_{flex}\in[\Upsilon_1, 100\%)$.}
		\label{fig3}
	\end{figure}
	
	To create two dimming ranges in an FEC-based system, a systematic FEC encoder should be applied to preserve the two probabilistic shapes of $\phi_{flex}$. Furthermore, the run-length of FEC codewords should be limited to avoid flickering at the system's optical clock. We solve these issues by employing the SPE with RLA frozen bit values. An SPC is specified by ($l$, $n$, $\digamma$), where $l$ is the code length, $n$ is the number of information bits encoded per codeword, and $\digamma$ is a set of $l$-$n$ indices which indicate locations of frozen bits, where $\digamma\subset \{1,2,3,..., l$\}, $\left| \digamma \right|$ $= l-n$. Given the SPC with code rate $R=n/l$, at a design signal-to-noise ratio (design-SNR), a construction algorithm over an additive white Gaussian noise (AWGN) channel selects $n$ best among $l$ bit channels for information bit indices, and the remaining $l-n$ channels for frozen bit indices. As the conventional setting, all 0's are assigned to frozen bits, which increases the 0's run-length of codewords. Hence, we introduce the RLA frozen bit values, which is addressed in Algorithm \ref{algor1}. As shown, the bit-channel metrics vector $z$, which is created by Arikan's Bhattacharyya bounds \cite{arikan2009channel}, is sorted in ascending order. For frozen bit indices, a bit sequence where the inverse values of bit-0 and bit-1 are placed one after another is assigned. By this setting, the free run of 0's or 1's is prevented. 
	
\LinesNumberedHidden
\begin{algorithm}[]
	\label{algor1}
	\resizebox{0.4\textwidth}{!}{%
		\begin{minipage}{0.53\textwidth}
			\KwData{Vector of bit-channel metrics $z \in {\mathbb{R}^l}$ generated from Bhattacharyya bounds \cite{arikan2009channel}.}
			\KwResult{ - Vector of frozen bit locations $\digamma$; $\left| \digamma \right|$ $= l-n$\\
				\hspace{13.3mm}      - Vector of RLA frozen bit values $y$;  $\left| y \right|$ = $l$ }
			\Begin{
				
				$t$ = sort($z$,'$ascending$') \tcp*{metrics ranking}
				\tcc{Best metrics for information bits, worst metrics for frozen bits}
				$\digamma$ = sort($t$[$n$ : $l]$, '$ascending$')\;
				\tcc{Sorting frozen bit locations in ascending order}
				\For{$i = 1:l-n $}{
					$j$ = $\digamma[i]$\;
					\If{$modulo(i,2) = 0$}{
						$y[j]$ = 0\;}
					\If{$modulo(i,2) = 1$}{$y[j]$ = 1\;}
				}
				\tcc{A sequence of 0,1,0,1... is assigned for frozen bit indices}
			}
		\end{minipage}
	}
	\caption{Method of setting RLA frozen bit values}
\end{algorithm}
	
	\begin{figure}[!t]
		\centerline{\includegraphics[width=0.95\columnwidth]{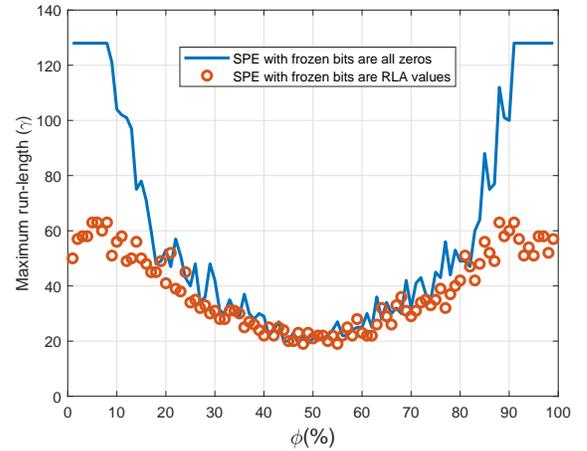}}
		\caption{Maximum run-length analysis of SPE codewords in two scenarios: all zeros or RLA values are assigned to frozen bit indices}
		\label{fig4}
	\end{figure}
	
	\begin{table*}[!htb]
		\caption{Comparison with hybrid modulation schemes introduced in IEEE P802.15.7m (TG7m) PHY-IV modes \cite{nguyen2018technical}}
		\label{table}
		\resizebox{0.45\textwidth}{!}{%
			\begin{minipage}{0.5\textwidth}
				\begin{threeparttable}
					\begin{tabular}{cccccc}
						\hline
						\begin{tabular}[c]{@{}c@{}}\textbf{Modulation/}\\\textbf{Coding}\end{tabular} & \textbf{RLL} & \textbf{Optical clock rate} & \textbf{FEC} & \textbf{Bit rate} & \begin{tabular}[c]{@{}c@{}}\textbf{Spectral efficiency}\\\textbf{(FEC excluded)}\end{tabular}\\ \hline
						Twinkle VPPM & N/A & 16 kHz & Reed-Solomon (15,11) & 4 kbps & 0.25 (bits/s)/Hz \\ \hline
						HS-PSK & \begin{tabular}[c]{@{}c@{}}1/2 code rate for S2-PSK;\\ None for DS8-PSK\end{tabular} & 10 kHz & \begin{tabular}[c]{@{}c@{}}Temporal error-correction;\\ Outer FEC with GF(16)\end{tabular} & 22 kbps & 0.1375 (bits/s)/Hz\tnote{$\dagger$}\\ \hline
						\begin{tabular}[c]{@{}c@{}}FDC \\ (\textit{Proposed})\end{tabular} & None & $>$12.6 kHz & \begin{tabular}[c]{@{}c@{}}Systematic polar code\\ (128,100)\end{tabular} & $>$ 11.47 kbps & 0.91 (bits/s)/Hz \\
						\bottomrule
					\end{tabular}
					\begin{tablenotes}
						\item[$\dagger$] HS-PSK: 2 light sources, 8 LEDs for each source. Average spectral efficiency = (Bit rate/(clock rate$\times$8$\times$2)).
					\end{tablenotes}
				\end{threeparttable}
			\end{minipage}
		}
	\end{table*}
	
	\section{Simulation results}
	It can be seen from Fig. \ref{fig2} that with $n = 100$, ${\Delta }_{FDC} = 0.09$ when $\phi_{flex}\in(0\%, \Upsilon_0]$ for logic 0 and $\phi_{flex}\in[\Upsilon_1, 100\%)$ for logic 1. On the other hand, ${\Delta}_{DC} > 0.1$ when $\phi<\Upsilon_0$ and $\phi>\Upsilon_1$. This can be explained by the fact that the codebook size of the FDC is much larger than that of the DC. In short, the FDC can produce two probabilistic shapes under the control of low-rate signal $v$ at very low rate loss. 
	
	Fig. \ref{fig3} shows the post-FEC-encoder probabilistic shapes. As can be seen from the figure, the two probabilistic ranges, $\omega\in(0\%, 50\%]$ and $\omega\in[50\%, 100\%)$, are not overlapped when the the FDC outputs are encoded by the SPE. In contrast, they overlap and stretch on the wide range, $\omega\in(10\%, 90\%)$, and tend to converge to the mid-range $\omega\in(40\%, 60\%)$ when a nonsystematic polar encoder (NSPE) is applied with the FDC. To enable low-rate communication from the two dimming ranges, the probabilistic ranges must not overlap each other to be discriminated by the low-speed camera on the receiver. Therefore, the SPE is selected as the FEC encoder in our system.
	
	We also evaluated the maximum run-length of codewords encoded by the SPE, which is denoted as $\gamma$. As shown in Fig. \ref{fig4}, when the RLA values are assigned to frozen bits, $\gamma = 63$ is reduced by a factor of 2.03 compared to when all-zeros are assigned to them. To avoid flicker, the LED's brightness should change within the maximum flickering time period (MFTP $\leq$ 5ms) \cite{rajagopal2012ieee}. Hence, with $\gamma$ = 63, our system guarantees the flicker-free operation with optical clock rates higher than 12.6 kHz without using an RLL code. 
	
	Table \ref{table} compares the proposed method with related work addressed in the image sensor communication PHY-IV modes added to TG7m \cite{nguyen2018technical}. The clock rate of our system (minimum is 12.6 kHz) is almost equivalent to that of related approaches. Given the fact that FEC codes with code rates $R\in (0.5, 0.8)$ have been introduced in PHY II modes of IEEE 802.15.7 and PHY-IV modes of TG7m \cite{rajagopal2012ieee,thieu2019optical}, we have evaluated our system with a polar code (128,100) with $R = 0.78$, where the reason for short code length $l=128$ is explained in section \ref{subsection_FDC}. With the FDC's rate loss ${\Delta }_{\phi_{flex}} = 0.09$, our system achieves a spectral efficiency 0.91 (bits/s)/Hz, which is better than systems based on hybrid modulation schemes.
	
	Without the support of RLL code (none-RLL), the CDR issues should be considered. Let $\beta$ denote the number of required image frames that the high-speed camera uses to synchronize the timing and decode the bit sequence on the receiver. In our previous work \cite{shiraki2021demodulation}, experimental results showed that when there is a clock difference between the transmitter and receiver, stable bit error rate (BER) performance is achieved with $60\leq\beta \leq100$. Fig. \ref{fig4} shows that the transmitted bit sequence in our system has maximum run-length $\gamma$ = 63. It can	be seen that by referring to $\gamma$, we can set $\gamma\leq\beta \leq100$ to guarantee the CDR in case of a long run of 1's or 0's appears in the bit sequence.
	
\begin{figure}[!t]
		\centerline{\includegraphics[width=1\columnwidth]{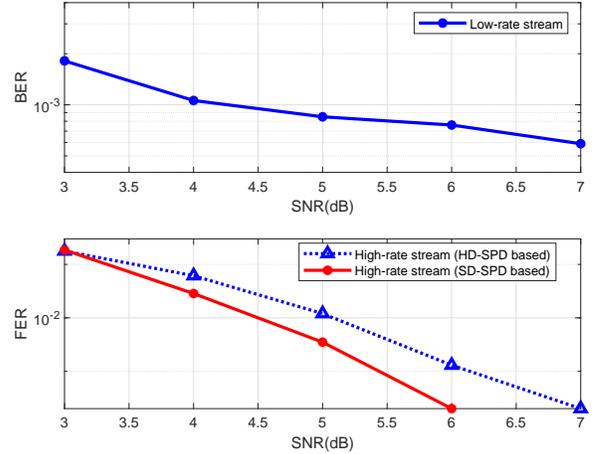}}
		\caption{BER and FER performance of the proposed system on low-rate and high-rate streams in an AWGN channel.}
		\label{fig5}
	\end{figure}

	The fact that two cameras use different mechanisms in demodulating the signals, we applied two different AWGN channel models to estimate the BER and frame error rate (FER) performances of the low-rate and high-rate data streams. Fig. \ref{fig5} shows the BER performance of the low-rate stream which is estimated from the overlap probability between the logic-0 and logic-1 probabilistic ranges of $\omega$ at different SNRs, without the support of error-correction scheme. Meanwhile, the FER of the high-rate stream is calculated when data are decoded by the SD-SPD and the inverse FDC. The BER performance of the low-rate stream can be improved with $\Upsilon_0 < 36\%$ and $\Upsilon_1 > 64\%$ if the system affords a higher code-rate loss. Fig. \ref{fig5} also shows that the coding gain of SD-SPD-based and HD-SPD-based systems differ by 1 dB when FER = 2E-4. In brief, due to the none-RLL feature, SD FEC decoders can be applied in our system to enhance system reliability. 
	
	\section{Conclusion}
	In this letter, we introduced a probabilistic shaping approach applied for RoI signaling. The desired flexible distribution of high-rate bit sequences are controlled by low-rate signal controls in the FDC. The SPC with the RLA frozen bit values minimizes the run-length and preserves probabilistic shapes created by the FDC. The low-speed camera decodes the low-rate data from two dimming ranges to identify the RoI, and then the high-speed camera decodes the high-rate data stream on the selected RoI. The proposed system achieves extremely good spectral efficiency with SD FEC decoding support, while the flicker mitigation is guaranteed without RLL code. 
	
	\bibliographystyle{IEEEtran}
	\bibliography{IEEEexample}

\begin{thebibliography}{1}
\providecommand{\url}[1]{#1}
\csname url@samestyle\endcsname
\providecommand{\newblock}{\relax}
\providecommand{\bibinfo}[2]{#2}
\providecommand{\BIBentrySTDinterwordspacing}{\spaceskip=0pt\relax}
\providecommand{\BIBentryALTinterwordstretchfactor}{4}
\providecommand{\BIBentryALTinterwordspacing}{\spaceskip=\fontdimen2\font plus
\BIBentryALTinterwordstretchfactor\fontdimen3\font minus
  \fontdimen4\font\relax}
\providecommand{\BIBforeignlanguage}[2]{{%
\expandafter\ifx\csname l@#1\endcsname\relax
\typeout{** WARNING: IEEEtran.bst: No hyphenation pattern has been}%
\typeout{** loaded for the language `#1'. Using the pattern for}%
\typeout{** the default language instead.}%
\else
\language=\csname l@#1\endcsname
\fi
#2}}
\providecommand{\BIBdecl}{\relax}
\BIBdecl

\bibitem{thieu2019optical}
M.~D. Thieu, T.~L. Pham, T.~Nguyen, and Y.~M. Jang, ``Optical-RoI-signaling for
  vehicular communications,'' \emph{IEEE Access}, vol.~7, pp. 69\,873--69\,891,
  2019.

\bibitem{nguyen2018technical}
T.~Nguyen, A.~Islam, T.~Yamazato, and Y.~M. Jang, ``Technical issues on IEEE
  802.15.7m image sensor communication standardization,'' \emph{IEEE
  Communications Magazine}, vol.~56, no.~2, pp. 213--218, 2018.

\bibitem{wang2017dimming}
H.~Wang and S.~Kim, ``Dimming control systems with polar codes in visible light
  communication,'' \emph{IEEE Photonics Technology Letters}, vol.~29, no.~19,
  pp. 1651--1654, 2017.

\bibitem{noh2015dimming}
J.~Noh, S.~Lee, J.~Kim, M.~Ju, and Y.~Park, ``A dimming controllable VPPM-based
  VLC system and its implementation,'' \emph{Optics Communications}, vol. 343,
  pp. 34--37, 2015.

\bibitem{rajagopal2012ieee}
S.~Rajagopal, R.~D. Roberts, and S.-K. Lim, ``IEEE 802.15.7 visible light
  communication: modulation schemes and dimming support,'' \emph{IEEE
  Communications Magazine}, vol.~50, no.~3, pp. 72--82, 2012.

\bibitem{wang2015soft}
H.~Wang and S.~Kim, ``Soft-input soft-output run-length limited decoding for
  visible light communication,'' \emph{IEEE Photonics Technology Letters},
  vol.~28, no.~3, pp. 225--228, 2015.

\bibitem{ramabadran1990coding}
T.~V. Ramabadran, ``A coding scheme for m-out-of-n codes,'' \emph{IEEE
  Transactions on Communications}, vol.~38, no.~8, pp. 1156--1163, 1990.

\bibitem{arikan2009channel}
E.~Arikan, ``Channel polarization: A method for constructing capacity-achieving
  codes for symmetric binary-input memoryless channels,'' \emph{IEEE
  Transactions on information Theory}, vol.~55, no.~7, pp. 3051--3073, 2009.

\bibitem{shiraki2021demodulation}
Y.~Shiraki, T.~G. Sato, Y.~Kamamoto, T.~Izumi, Y.~Nakahara, K.~Kondo, and
  T.~Moriya, ``A demodulation method using a gaussian mixture model for
  unsynchronous optical camera communication with on-off keying,''
  \emph{Journal of Lightwave Technology}, vol.~39, no.~6, pp. 1742--1755, 2021.

\end{thebibliography}
	
	\appendices
	
\end{document}